\documentclass[aps,prl,twocolumn,floatfix]{revtex4}

\usepackage{amsmath,amssymb,dsfont}
\usepackage{color}
\usepackage[pdftex]{hyperref,graphicx}
\hypersetup{colorlinks = true, urlcolor = blue, linkcolor = blue, citecolor = blue}

\providecommand{\tgamma}{\ensuremath{\tilde{\gamma}}}

\begin{document}
\title{Generalized Jackiw-Rebbi Model and Topological Classification of \\ Free Fermion Insulators}
\author{O. Nganba Meetei and Archana Anandakrishnan}
\affiliation{Department of Physics, The Ohio State University, Columbus, Ohio 43210, USA} 
\begin{abstract}
We present a new perspective to the classification of topological phases in free fermion insulators by generalizing the Jackiw-Rebbi model to arbitrary dimensions. We show that a generalized Jackiw-Rebbi model where the Dirac mass ($m$) satisfies $m(x)=-m(-x)$ is invariant under a parity transformation ($P$) that relates the $x>0$ half to the $x<0$ half. Determining the form of $P$ gives rise to a Clifford algebra that has been shown to give a complete topological classification of free fermion insulators. Gapless edge states are a natural consequence of our construction and their topological nature can be understood from the fact that all gapless edge states at a given interface transform similarly under $P$ (all odd or all even). A naive non-topological model for states confined to the interface will allow both even and odd states.  
\end{abstract}
\date{\today}
\maketitle

In the Landau paradigm, phase transitions are characterized by spontaneously broken symmetries. However, the discovery of integer quantum Hall states \cite{klitzing_1980} showed that there are states of matter which are clearly distinct from each other but do not break any symmetry and, therefore, require a topological classification \cite{TKNN_1982}. More recently, the theoretical prediction \cite{kane_mele_2005,bernevig_quantum_2006,fu_prl_2007,moore_2007,roy_2009} and subsequent experimental discovery \cite{molenkamp_2007,hasan_2009} of time reversal invariant topological insulators have added a new class of free fermion insulators where gapless edge states are protected by some symmetry of the system. Such topological insulators are smoothly connected to trivial insulators once the symmetry is broken and they come under a broad family called Symmetry Protected Topological (SPT) insulators. 

SPT insulators are a very active topic of research, with a rapidly growing list of predicted SPT insulators \cite{hasan_colloquium_2010,qi_rmp_2011,kargarian_2011,hsieh_2012,chen_2012}. The need to bring order to this zoo of SPT insulators has renewed interest in topological classification of phases. Classification of SPT phases depend crucially on the symmetries and dimensionality of the system. In the case of effectively free fermion systems, two independent schemes have been developed recently: K-theory approach by Kitaev \cite{kitaev_2009} and nonlinear $\sigma$ model for disordered fermions by Schnyder {\it et al.} \cite{schnyder_2008}. Both approaches provide a comprehensive classification of all topological insulators and superconductors. 

In this paper, we present a new perspective to the classification of free fermion SPT insulators by reconnecting them to generalizations of the Jackiw-Rebbi model \cite{jackiw_rebbi_1976} to arbitrary dimensions. Historically, the Jackiw-Rebbi model is significant as one the earliest theoretical  description of a topological edge state. There has been hints of a broader connection between the Jackiw-Rebbi model and topological insulators \cite{dh_lee_2007,nishida_2010, ho_majorana_2012}, but a systematic link has not been established yet. Here, we show a rigorous mapping of all classes of free fermion SPT insulators to a suitable Jackiw-Rebbi model. Its simplicity lends insight into the problem and it is by construction well suited for determining the nature of topologically protected edge states.

The main results presented in this paper are: 1) A generalized Jackiw-Rebbi model where the Dirac mass ($m$) satisfies $m(x)=-m(-x)$ is invariant under a parity transformation $P$ which relates the $x>0$ half to the $x<0$ half. Determining the form of $P$ yields a Clifford algebra identical to Kitaev's approach. So, for $d \geq 1$ the classification of $P$ leads to a complete topological classification of free fermion insulators.  2) The $m>0$ and $m<0$ halves of the generalized Jackiw-Rebbi model correspond to different topological phases in the same symmetry class. In other words, the interface between any two topologically different insulators in the same symmetry class can be represented by a suitable Jackiw-Rebbi model. 3) All symmetry protected gapless edge states at the interface transform similarly under $P$ (all odd or all even). This explains their topological nature because only half of the possible states from a naive non-topological model for the interface are allowed as edge states. Once the form of $P$ is fixed, the structure of edge states ({\it e.g.} spin-momentum locking) is completely determined. 


\begin{figure}[t]
\includegraphics[width=6.5cm]{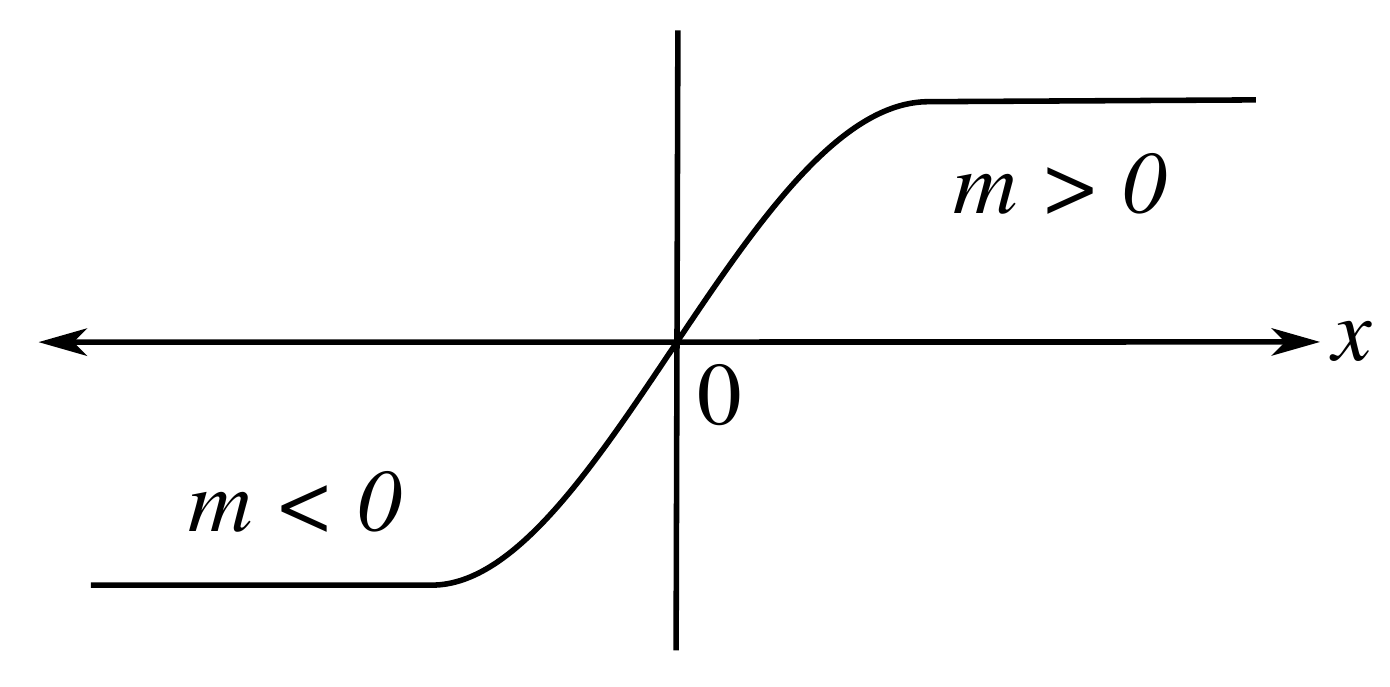}
\caption{Schematic of the generalized Jackiw-Rebbi model in $d$-dimensions where the Dirac mass term ($m$) changes sign across the $x=0$ interface, and satisfy $m(x)=-m(-x)$. $m$ is independent of the remaining $d-1$ dimensions.}
\label{fig:schematic}
\end{figure}

\emph{Generalized Jackiw-Rebbi Model}: The original model proposed by Jackiw and Rebbi \cite{jackiw_rebbi_1976} describes a one dimensional system where Dirac fermions are coupled to a soliton field. They showed that a kink in the soliton field which is equivalent to the Dirac mass ($m$) changing sign across an interface leads to a topologically protected zero energy mode localized at the interface. Re-interpreting the $m>0$ and the $m<0$ halves as different topological phases, the zero energy mode can be understood as the edge state one expects at the interface of two topologically different states of matter. 

Here we generalize the Jackiw-Rebbi model to arbitrary dimensions $d$, where $m$ changes sign at the $x=0$ interface as shown in the schematic diagram in Fig.~\ref{fig:schematic}. $m$ is independent of the remaining $(d-1)$ physical dimensions. We require that $m(x)=-m(-x)$ and choose $m(x)>0$ for $x>0$. The Dirac Hamiltonian describing such a system has the form 
\begin{equation}\label{eq:Dirac_Ham}
\begin{split}
   H(x) &= -i\gamma^\mu\partial_\mu + m(x)\gamma^0 \\
   m(x) &= \left\{
	 \begin{array}{rr}
		\lvert m(x) \rvert , \; x>0 \\
		-\lvert m(x) \rvert , \; x<0 
	 \end{array}	\right.
\end{split}
\end{equation}
where the $\gamma$ matrices are in general complex Hermitian matrices and satisfy
\begin{eqnarray}  \label{eq:hermitian_gamma}
   (\gamma^0)^2=\mathds{1} \text{  and  } \gamma^\mu\gamma^0 = -\gamma^0 \gamma^\mu, \;\; \mu = 1,.., d \nonumber \\
   (\gamma^\mu)^2=\mathds{1} \text{  and  } \gamma^\mu\gamma^\nu = -\gamma^\nu \gamma^\mu \;\; \forall \; \mu\neq\nu 
\end{eqnarray}  
In order to describe insulators with time reversal symmetry and superconductors, the Majorana representation becomes important where the matrices are real and the Dirac Hamiltonian can be written as
\begin{equation}\label{eq:Majorana_Ham}
  H(x) = -i\left( \tilde{\gamma}^\mu\partial_\mu + m(x)\tilde{\gamma}^0 \right)
\end{equation}
where 
\begin{eqnarray}  \label{eq:real_gamma}
   (\tgamma^0)^2=-\mathds{1} \text{  and  } \tgamma^\mu\tgamma^0 = -\tgamma^0 \tgamma^\mu, \;\; \mu = 1,.., d \nonumber \\
   (\tgamma^\mu)^2=\mathds{1} \text{  and  } \tgamma^\mu\tgamma^\nu = -\tgamma^\nu \tgamma^\mu \;\; \forall \; \mu\neq\nu 
\end{eqnarray}  

$\tilde{\gamma}^0$ is a real anti-symmetric matrix while the rest are real symmetric matrices. In the rest of the paper, $\lbrace \tilde{\gamma}^\mu\rbrace$ will indicate the real case while $\lbrace\gamma^\mu\rbrace$ will mean the complex case. Note that the real Majorana representation and the complex Dirac fermion representation are related by a basis transformation.


We focus on Dirac Hamiltonians because all gapped free fermion systems close to a topological phase transition can be described by the appropriate Dirac Hamiltonian in the continuum limit. The structure of the Dirac spinor is related to degrees of freedom like spin and band index relevant for the system. 

\emph{Parity Transformation}: We now show that we can always define a parity transformation $P:x\rightarrow -x$ under which the Dirac Hamiltonians in Eq.~\ref{eq:Dirac_Ham} and Eq.~\ref{eq:Majorana_Ham} are invariant. The interface at $x=0$ maps onto itself under this operator. For the complex case in Eq.~\ref{eq:Dirac_Ham}, consider the form $P=iO\gamma^1 X$ where $X: x\rightarrow -x$ acts on the co-ordinate space  and $O\gamma^1$ acts on the spinor space. For $P$ to a good symmetry of the model, we require that 
\begin{equation}
P H P = H \quad \text{and} \quad P^2 = \mathds{1}  
\end{equation} 
which yields the following constraints on $O$ 
\begin{equation}
\begin{split}
& O^2 = \mathds{1}\; ; \quad  O\gamma^0 = \gamma^0 O \\
& O\gamma^\mu = -\gamma^\mu O \;\; \text{for}  \;\; \mu=1,...,d
\end{split}
\end{equation}
So, without loss of generality, we can choose $O=\gamma^0$ and the parity transformation which leaves the system invariant is given by  
\begin{equation} \label{eq:complex_P}
P = i \gamma^0\gamma^1 X
\end{equation}
For the real case, a similar analysis shows that the parity operator has the form
\begin{equation} \label{eq:real_P}
P=\tilde{\gamma}^0\tilde{\gamma}^1 X
\end{equation}

\emph{Topological Classes}: In Kitaev's K-theory approach to topological classification of free fermion insulators, topological insulators with no additional symmetries are described by the complex Dirac equation in Eq.~\ref{eq:Dirac_Ham}. The matrices $\gamma^1,..,\gamma^d$ in Eq.~\ref{eq:hermitian_gamma} form a representation of the complex Clifford algebra $\mathcal{C}l_d(\mathds{C})$ and the goal is to find an additional Clifford generator $\gamma^0$ which acts as the mass term and opens a gap. The space for $\gamma^0$ is $C_{(d \text{ mod } 2)}$ (see Table 2 of Ref.~[\citenum{kitaev_2009}]) and the topological classification  is given by the zeroth order homotopy group $\pi_0(C_{(d \text{ mod } 2)})$ which is $\mathds{Z}$ for even $d$ and $0$ for odd $d$. As shown above in Eq.~\ref{eq:complex_P}, finding $\gamma^0$ for a given set of $\gamma^\mu \; (\mu=1,...,d)$ is equivalent to finding the parity operator $P$. So, for the complex case, there is a one to one mapping between the topological classes and a generalized Jackiw-Rebbi model. 

We now turn to the real Majorana representation in Eq.~\ref{eq:Majorana_Ham} which is useful for classifying insulators with additional symmetries like time-reversal ($T$) and/or charge conjugation $(C)$. In general, let $A_i, \; i=1,..,p+q$ denote the symmetry operators of the system such that $A_i^2 = \mathds{1}$ for $i=1,..,p$ and $A^2_{p+i} = -\mathds{1}$ for $i=1,..,q$.  The matrix representations of the symmetry operators can be chosen so that $A_iA_j=-A_jA_i \;\; \forall \; i\neq j$ and $A_i\tgamma^\mu = -\tgamma^\mu A_i$ for $\mu =1,..,d$. Combining these requirements with the properties of $\tgamma$ matrices in Eq.~\ref{eq:real_gamma}, it is clear that $\lbrace A_i\rbrace , \; i=1,..,p+q$ and $\lbrace \tgamma^\mu \rbrace, \; \mu =1,..,d$ form a representation of the real Clifford algebra $\mathcal{C}l^q_{p+d}(\mathds{R})$ with $p+d$ positive generators and $q$ negative generators. The space of insulators with symmetries $\lbrace A_i\rbrace$ in $d$-dimensions is then described by the space of the mass term $\tgamma^0$ which is an additional negative Clifford generator. Table 3 of Ref.~[\citenum{kitaev_2009}] shows that the space of $\tgamma^0$ is $R^q_{p+d} \simeq R_{(q-p-d+2) \text{ mod }8}$ and the topological classification is given by $\pi_0(R_{(q-p-d+2) \text{ mod }8})$. Just as in the complex case, the task of finding $\tgamma^0$ given $A_i,..,A_{p+q},\tgamma^1,..,\tgamma^d$ is equivalent to finding the parity operator $P$ defined in Eq.~\ref{eq:real_P}. So, the topological classification of symmetry protected free fermion insulators is equivalent to the classification of the appropriately represented $P$ operator. 

By construction, the Jackiw-Rebbi model is defined only for $d\geq 1$ which leaves out the $d=0$ case. While the $d=0$ case is interesting, it doesn't give rise to any edge states. The focus of our paper is to develop a simple and yet powerful theory for the interface between different topological states. 

\emph{Edge States}: The structure of the Jackiw-Rebbi model is ideal for determining the properties of the edge states. First, we show that our construction guarantees the existence of gapless edge states. Then, we show that the edge states are topological in nature which implies that the $m>0$ and $m<0$ halves represent different topological states in the same symmetry class.  

For brevity, we solve for the edge states in the complex case only. The steps are identical in the real Majorana representation.   Consider the following ansatz for the edge states
\begin{equation}
\psi_{\mathbf k_\perp}(x, {\mathbf r_\perp}) = \phi_{\mathbf k_\perp} \; e^{-\int^x_0 m(x)\text{d}x}e^{-i{\mathbf k_\perp \cdot \mathbf{r}_\perp}}
\end{equation} 
where ${\mathbf k_\perp}$ and ${\mathbf r_\perp}$ are the momentum and position vectors perpendicular to x. $\phi_{\mathbf k_\perp}$ is a spinor whose dimension is determined by the band and/or spin index. Plugging this ansatz into the Dirac equation in Eq.~\ref{eq:Dirac_Ham}, we get
\begin{equation} \label{eq:edge1}
\left( i m(x) \gamma^x + k_i \gamma^i + m(x) \gamma^0 \right) \psi_{\mathbf k_\perp} = E \psi_{\mathbf k_\perp}
\end{equation}
For the zero energy mode, we set $E=0$ and $\mathbf k_\perp =0$. After multiplying Eq.~\ref{eq:edge1} on the left by $\gamma^0$ and using the definition of $P$ in Eq.~\ref{eq:complex_P}, we get
\begin{equation}
   m(x) \left( P + \mathds{1} \right) \psi_0 = 0
\end{equation} 
It is clear that $\psi_0$ must be an eigenstate of $P$ with eigenvalue -1. Since $P$ is a good symmetry of the Hamiltonian, all other $E \neq 0$ edges states which are smoothly connected to $\psi_0$ as a function of $\mathbf k_\perp$ must also be odd eigenstates of $P$. The even eigenstates of $P$ are not allowed as gapless edge states. Note that if we had chosen $P=-i \gamma^0\gamma^1 X$ which differs from the definition of $P$ in Eq.~\ref{eq:complex_P} by a (-) sign, only the even eigenstates of $P$ would be permissible. 

Let us compare the gapless edge states obtained from the generalized Jackiw-Rebbi model to the states of a massless Dirac Hamiltonian. If we set $x=0$ and $m(0)=0$ in Eq.~\ref{eq:edge1}, the massless Dirac Hamiltonian in $(d-1)$ dimensions is the naive Hamiltonian for the edge states. 
\begin{equation}
	H_{Edge} = -i\gamma^\mu\partial_\mu \quad i = 2,..,d
\end{equation}
Using the properties of $\gamma$ matrices in Eq.~\ref{eq:hermitian_gamma} and the definition of $P$ in Eq.~\ref{eq:complex_P}, it is easy to show that $[P,H_{Edge}]=0$, and we can find common eigenstates of $P$ and $H_{Edge}$. The requirement that only odd eigenstates of $P$ are allowed as edge states projects out half of the eigenstates of $H_{Edge}$. This illustrates the topological nature of the gapless edge modes. The $m>0$ half is topologically different from the $m<0$ half. In other words, the interface between any two topologically inequivalent free fermion insulators belonging to the same symmetry class can be represented by a suitable generalized Jackiw-Rebbi model. This general result provides us with a convenient method for determining the nature of topological edge states as demonstrated below with some simple examples.

The fact that half of the eigenstates of $H_{Edge}$ are not allowed as edge states does not violate any fermion doubling theorem \cite{nielsen_ninomiya_1981}. One can imagine periodic boundary condition along $x$ with periodicity $2L$. Then, another kink in the mass term will occur at $x=L$ which is the opposite of the kink at $x=0$. The other half of eigenstates of the interface Hamiltonian $H_{Edge}$ are localized at the $x=L$ interface. 

\emph{Examples}: We now demonstrate the usefulness of our approach by using two well known examples: 1) Integer Quantum Hall Effect (IQHE) and 2) Quantum Spin Hall Effect (QSHE). We construct the lowest dimensional matrix representations of each and indicate how it can be generalized to higher dimensional representations. 

\begin{figure}[t!]
\includegraphics[width=7cm]{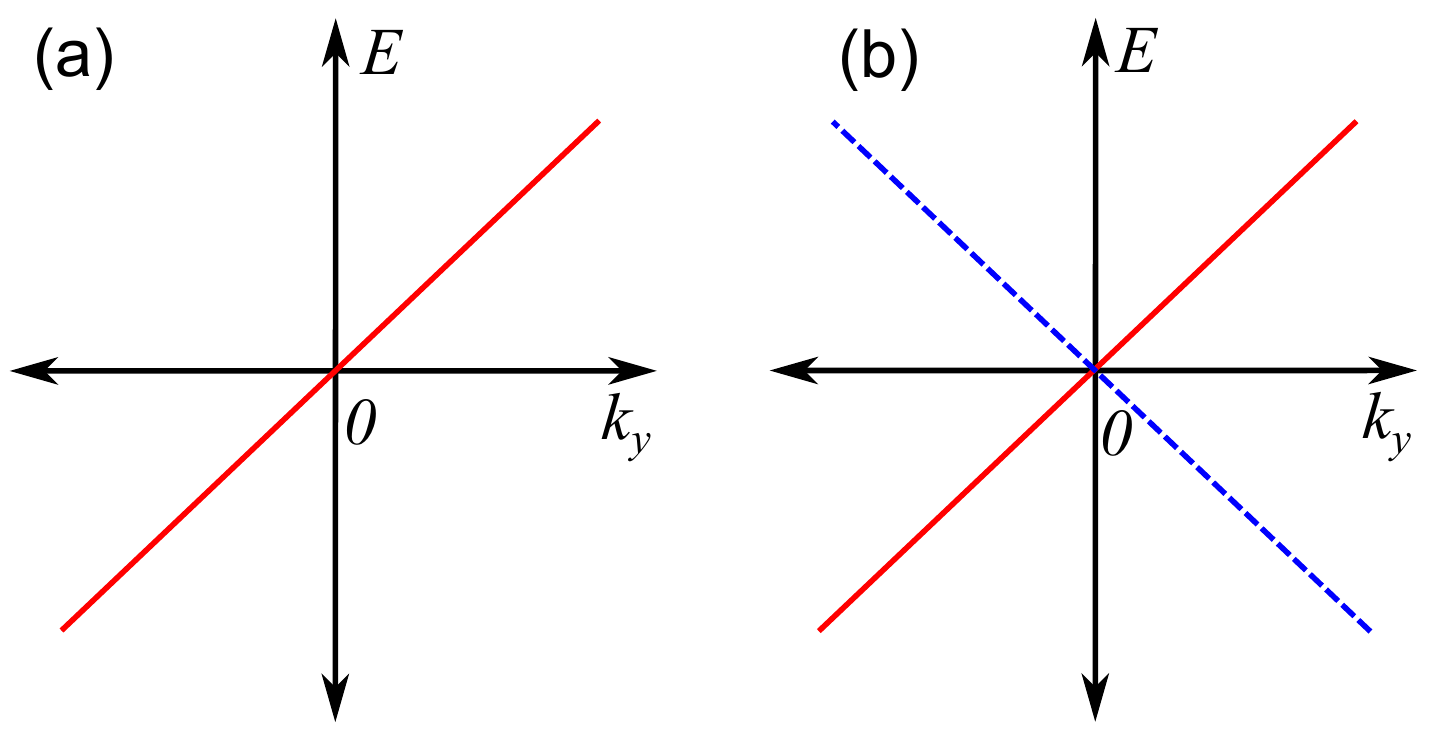}
\caption{Schematic of (a) the chiral edge state of an Integer Quantum Hall state and (b) the helical edge states of a Quantum Spin Hall insulator. Red (solid) line indicates spin down state and blue (dashed) line indicates spin up states.  }
\label{fig:edge_states}
\end{figure}

IQHE occurs in $d=2$ and there is no time reversal symmetry. So it belongs to the complex representation. The only symmetry is charge conservation or $U(1)$ symmetry which is trivially satisfied by the Dirac equation. In the lowest dimensional representation, we can choose $\gamma^1=\sigma^x$ and $\gamma^2=\sigma^y$. For this choice of $\gamma^1$ and $\gamma^2$, the only allowed form for the mass term is $\gamma^0=\sigma^z$. Here $\sigma$'s are the Pauli matrices. Then the Hamiltonian for the gapless edge states and the parity operator are given by
\begin{equation}
   H_{Edge}= -i\sigma^y\partial_y  \qquad P=i\sigma^z\sigma^xX = -\sigma^yX
\end{equation}
where we have set $v_f=1$. Restricting the allowed edge states to the odd eigenstates of $P$ selectively picks out the positive eigenstates of $\sigma^y$ or the right moving eigenstates of $H_{Edge}$. These are the chiral edge states of the integer quantum Hall states which emerge naturally from our construction (see Fig.~\ref{fig:edge_states}(a)).

Generalizations to higher dimensional representations can be achieved by constructing block diagonal matrices. Keeping in mind that K-theory allows supplementing the Hamiltonian by a trivial piece, we can choose the following form for the $\gamma$ matrices:
\begin{equation} \label{eq:IQHE}
\begin{split}
\gamma^1 &= I_{n\times n}\otimes \sigma^x \; ;\qquad \gamma^2 = I_{n\times n}\otimes\sigma^y \\
\gamma^0 &= \left( \begin{array}{cc} I_{l\times l} & 0 \\ 0 & -I_{m\times m} \end{array} \right)\otimes \sigma^z \quad ; \;\; l+m=n
\end{split}
\end{equation}
where $I_{n \times n}$ is the $n\times n$ identity matrix. This gives rise to $l$ right moving and $m$ left moving edge states.  Note that the Pauli matrices along with the identity matrix exhausts the space $2\times 2$ Hermitian matrices. So, the most general unitary transformation that leaves both $\gamma^1$ and $\gamma^2$ in Eq.~\ref{eq:IQHE} invariant has the form $U\otimes I_{2\times 2}$ where $U$ is any $n \times n $ unitary matrix. $\gamma^0$ is not invariant under this transformation and becomes
\begin{equation}
 \gamma^0=U \left( \begin{array}{cc} I_{l\times l} & 0 \\ 0 & -I_{m\times m} \end{array} \right) U^\dagger\otimes \sigma^z
\end{equation} 
Here, the unitary transformation only affects the first factor in $\gamma^0$ and does not change its eigenvalues. Therefore, the difference in the number of right and left moving edge states ($l-m\in \mathds{Z}$) is a topologically invariant quantity. 

Moving on to the next example, QSHE requires time reversal symmetry in addition to the $U(1)$ symmetry. We consider the BHZ Hamiltonian for HgTe/CdTe \cite{bernevig_quantum_2006} but retain terms only upto linear order in $\mathbf k$. Replacing $k_\mu$ by $-i\partial_\mu$ gives the required Dirac equation with $\gamma^1=\sigma^z\otimes \sigma^x$, $\gamma^2=-I_{2\times 2}\otimes \sigma^y$ and $\gamma^0=I_{2\times 2}\otimes \sigma^z$. The top $2\times 2$ block corresponds to the conduction and valence band spin up states while the bottom $2\times 2$ block describes the corresponding spin down states. The two blocks are related by time reversal symmetry. Like in the case of IQHE, we set $v_f=1$ without loss of generality. Then, the Dirac Hamiltonian for the edge states and the parity operator have the form
\begin{equation}
\begin{split} 
H_{Edge} &=-i\left( \begin{array}{c c} -\sigma^y & 0 \\ 0 & -\sigma^y \end{array} \right) \partial_y \\ 
P &= \left( \begin{array}{c c} -\sigma^y & 0 \\ 0 & \sigma^y \end{array} \right)X
\end{split}
\end{equation}
Since both the edge Hamiltonian and the parity operator are block diagonal, we can look at each spin species separately. For the spin down sector or the lower $2\times 2$ block, the odd eigenstate of $P$ means the negative eigenstate of $\sigma^y$ or, equivalently, the right moving eigenstate of $H_{Edge}$. Similarly for the spin up sector, the odd eigenstate of $P$ picks out the left moving eigenstate of $H_{Edge}$. So, we have gapless edge modes consisting of right moving spin down states and left moving spin up states, as shown schematically in Fig.~\ref{fig:edge_states}(b). Such helical edge states have been predicted for QSHE \cite{qi_rmp_2011} and it appears in a transparent way in our analysis.    

Higher dimensional representations can be constructed in a block diagonal fashion, in a way very similar to the case of IQHE. 

\begin{equation} \label{eq:QSHE}
\begin{split}
& \gamma^1=I_{n\times n}\otimes\sigma^z\otimes \sigma^x  \quad \gamma^2= I_{n\times n}\otimes(-I_{2\times 2})\otimes \sigma^y \\
&\gamma^0 = \left( \begin{array}{c c} I_{l\times l} & 0 \\ 0 & -I_{m\times m} \end{array} \right)\otimes I_{2\times 2}\otimes \sigma^z \; ; \quad  l+m=n
\end{split}
\end{equation}
If we focus on one spin sector, it consists of $l$ negative and $m$ positive helicity states. While the edge states are always helical, the sign of helicity is not a robust quantity. Both $\gamma^1$ and $\gamma^2$ in Eq.~\ref{eq:QSHE} are invariant under a unitary transformation by $\mathcal{U}=U\otimes exp(i \theta \, \sigma^y\otimes\sigma^y)$. For $\theta=\pi/2$, $\gamma^0$ transforms as 
\begin{equation}
\mathcal{U}\gamma^0 \mathcal{U}^\dagger =  U \left( \begin{array}{c c} -I_{l\times l} & 0 \\ 0 & I_{m\times m} \end{array} \right)U^\dagger \otimes I_{2\times 2}\otimes \sigma^z
\end{equation}
which exchanges the number of positive and negative helicity edge states. The topologically invariant quantity in this case is whether the number of edge states is odd or even ( $(l+m) \text {mod}\; 2 \in \mathds{Z}_2$).

In the current analysis for QSHE, we have used the complex representation. It can be equivalently done in the real basis by choosing the Majorana operator to be $\eta_{1\sigma}(\mathbf k)=c_\sigma(\mathbf k)+c_\sigma^\dagger(\mathbf k)$ and $\eta_{2\sigma}(\mathbf k)=-i(c_\sigma(\mathbf k)-c_\sigma^\dagger(\mathbf k))$. The simplest representation of real $\tgamma$'s consists of $8\times 8$ matrices. While the real representation is important for classification, it is not crucial for determining the edge states. In fact, the lower dimensional complex representation is much simpler to deal with and physical interpretation of the edge states is easier.  

In conclusion, we have established that there is a deep connection between the symmetry protected free fermion topological insulators and generalized versions of the Jackiw-Rebbi model. Every pair of topologically different insulators belonging to the same symmetry class can be mapped to a Jackiw-Rebbi model  where $m(x)=-m(-x)$.  We have defined a parity operator ($P$) which maps the $x<0$ half to the $x>0$ half and is a good symmetry of the model. The topological classification of symmetry protected free fermion insulators in $d\geq 1$ is equivalent to the classification of the $P$ operator. The simplicity of the model provides insights into the topological nature of the edge states. Our analysis yields a general scheme for determining the structure of gapless edge states. One simply needs to find the common eigenstates of the massless Dirac Hamiltonian for the edge and $P$, and keep only the odd eigenstates of $P$. While $P$ plays a crucial role in our construction, any smooth deformation which breaks $P$-symmetry without closing the bulk gap does not change the topology of the states. In this paper, we have illustrated the usefulness of our approach by using only two of the well known examples (IQHE and QSHE). We hope to extend the analysis to a larger class of symmetry protected topological insulators in the future. 

\emph{Acknowledgments}: We thank Nandini Trivedi, Mohit Randeria, Stuart Raby and Mehdi Kargarian for useful discussions. Our research was supported by Center for Emergent Materials funded by NSF
MRSEC DMR-0820414 (ONM) and the Ohio State University Presidential Fellowship (AA).

\bibliography{references}
\bibliographystyle{apsrev4-1}

\end{document}